\documentclass{elsart5p}

 \usepackage{graphicx}

\usepackage{amssymb}
\journal{Physica C}

\begin{document}

\begin{frontmatter}




\title{Observation of Vortex Coalescence, Vortex Chains and Crossing Vortices in the Anisotropic Spin-Triplet Superconductor Sr$_{2}$RuO$_{4}$}

 \author[label1]{K. Hasselbach\corauthref{cor}},
 \corauth[cor]{Corresponding author.}
 \ead{klaus.hasselbach@grenoble.cnrs.fr}
 \author[label1]{V.O. Dolocan\thanksref{now}},
 \thanks[now]{Present address: Max Plank Institute for Chemical Physics of Solids, 40 N\"{o}thnitzerStr., 01187 Dresden, Germany}
 \ead{dolocan@cpfs.mpg.de}
 \author[label1]{P. Lejay}
 \author[label2]{Dominique Mailly}
 
\address[label1]{CRTBT-CNRS, BP 166X, 38042 Grenoble, France}
\address[label2]{LPN-CNRS, Route de Nozay, 91460 Marcoussis, France}

\begin{abstract}

 Scanning $\mu$SQUID force microscopy is used to study magnetic flux structures in single crystals of the layered spin
triplet superconductor Sr$_{2}$RuO$_{4}$.  Images of  the magnetic flux configuration above the $\vec{a}\vec{b}$-face of the cleaved crystal are acquired, mostly after field-cooling the sample. For low applied magnetic fields, individual vortices  are observed, each carrying a single quantum of flux. Above 1 gauss, coalescence of vortices is discovered. The coalescing vortices may indicate the presence of domains of a chiral order parameter.

When the applied field is tilted from the $\vec{c}$-axis, we observe a gradual transition from vortex domains to vortex chains.  The  in-plane component of the applied magnetic field transforms the vortex domains to vortex chains by aligning them along the field direction.This behavior and the inter-chain distance varies in qualitative agreement with the Ginzburg Landau theory of anisotropic 3D superconductors. The effective mass anisotropy of Sr$_{2}$RuO$_{4}$, $\gamma$=20, is the highest observed in three dimensional superconductors. 

When the applied field is closely in plane, the vortex form flux channels confined between the crystal-layers. Residual Abrikosov vortices are pinned preferentially on these channels. Thus the in-plane vortices are decorated by crossing Abrikosov vortices: two vortex orientations are apparent
simultaneously, one along the layers and the other perpendicular to
the layers. 

\end{abstract}

\end{frontmatter}

\section{Introduction}
\label{Introduction}
 Sr$_{2}$RuO$_{4}$ is a tetragonal, layered perovskite superconductor with a superconducting transition temperature (T$_{c}$) of 1.5 K \cite{Mae94}.  Sr$_{2}$RuO$_{4}$ has been a subject of intensive interest in recent years because of the theoretical suggestion \cite{Ric95,Bas96} that  Sr$_{2}$RuO$_{4}$ is an odd-parity, spin-triplet superconductor. Abundant experimental evidence supporting the theoretical prediction has been obtained, as summarized recently \cite{Mac03,Mae01}. A very recent phase-sensitive experiment \cite{Nel04} has established the odd-parity pairing symmetry in  Sr$_{2}$RuO$_{4}$ by measuring the quantum interference pattern in Superconducting Quantum Interference Devices (SQUIDs) consisting of  Sr$_{2}$RuO$_{4}$ and Au$_{0.5}$In$_{0.5}$, an s-wave superconductor. In addition, muon spin rotation ( $\mu$SR ) experiments have revealed \cite{Luk98}  the presence of spontaneous currents in the superconducting  Sr$_{2}$RuO$_{4}$, indicating the breaking of time reversal symmetry (TRS) below T$_{c}$. The TRS breaking implies that the Cooper pair has an internal orbital moment (chirality) giving rise to a superconducting order parameter with multiple components. 

The crystal structure and the thermodynamic properties of the
superconductor restrain the choice of the order parameter.
 The orbital component of the order parameter of the form, (p$_{x}$$\pm$ ip$_{y}$)  is compatible with most experiments. The two possible realisations of the superconducting order parameter, p$_{x}$+ ip$_{y}$  (p+) and p$_{x}$- ip$_{y}$  (p-), represent two possible chiral states \cite{Min_book} which are energetically degenerate. Consequently the presence of domains  is expected in which the Cooper pairs posses different orbital angular momenta. Building on this form of the order parameter, the magnetization processes were explored by numerical simulations \cite{Ich02,Ich05}. In a magnetic field the degeneracy between p+ and p- domains is lifted, favoring a domain with the Copper pair orbital moment aligned with the field. The favored domains will have a higher critical field H$_{c2}$, and a lower H$_{c1}$. Consequently, vortices will first appear in these domains. At the interface between the domains of opposite chiralities, walls will form \cite{Sig99}. The presence of domain walls has considerable influence on the vortex motion and pinning. For example, flux penetration should take place preferentially along the domain walls. These walls will act as preferential pinning sites for vortices. Vortices at these sites could decompose into fractional vortices decorating the domain walls. In analogy with the case of superfluid $^{3}$He-A decorated domain walls are called vortex sheets \cite{Mat04,Parts_94}.

The magnetic properties of superconductors depend strongly on their
crystalline and electronic anisotropy. The general theoretical approach on vortex
matter is based on the anisotropic Ginzburg-Landau (GL) theory. There the
anisotropy is expressed in terms of the effective mass of the electron. For
layered anisotropic superconductors, the out of plane effective mass
m$_{c}$ is much larger than the in plane effective masses
(m$_{c}$$>>$m$_{ab}$). To describe this anisotropy the parameter
$\gamma$=(m$_{c}$/m$_{ab}$)$^{1/2}$=$\lambda_{c}/\lambda_{ab}$\cite{Cle98}
is used. For example in NbSe$_{2}$ $\gamma$=3.3, in YBCO$\gamma$=5-8
and in BSCCO $\gamma$ is higher than 150, $\gamma$ being dependent  on the
oygen doping of the high T$_{c}$ superconductors.
Sr$_{2}$RuO$_{4}$ has a also a layered structure, the RuO$_{2}$
planes are separated by 12.74 {\AA} and has highly anisotropic
properties\cite{Mac03}. Sr$_{2}$RuO$_{4}$ has a $\gamma$ value of 20
situating it between YBCO and BSCCO on the anisotropy scale. We
expect Sr$_{2}$RuO$_{4}$ to act more like a 3D superconductor as the
c-axis parameter is 3 times smaller than the coherence length
$\xi_{c}$. The Ginzburg-Landau parameter $\kappa$ = $\lambda/\xi$ is
around 2.3 when the magnetic field is applied along the c-axis direction and 46 for the
in-plane direction. The physical properties of
Sr$_{2}$RuO$_{4}$ are very rich due to its unconventional mecanism of superconductivity and its anisotropy.
Scanning magnetic probe microscopy\cite{Gri01,Dolo05}.
 is a means of choice to study this interplay.

\section{$\mu$ SQUID Force microscopy and crystals }
\label{microscope}

\begin{figure}[!t]
 \includegraphics[width=7cm]{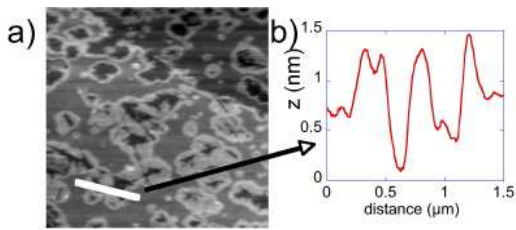}\\
 \caption{\label{Figure_1_Dresden_surface}(a)AFM Image of the surface of cleaved Sr$_{2}$RuO$_{4}$(b)Line scan as indicated, the roughness is of the order of half the unitcell.} \end{figure}

We use for magnetic imaging a high resolution scanning $\mu$SQUID
microscope (S$\mu$SM)\cite{Veau02} working in a dilution
refrigerator. The S$\mu$SM has an aluminum $\mu$SQUID as pickup loop
of 1.2$\mu$m diameter. The critical current of the $\mu$SQUID is a
periodic function of the magnetic flux emerging perpendicularly from
the sample surface. The images shown are maps of the critical
current value of the $\mu$SQUID while it scans the surface.
The magnetic fields are applied by a solenoid and a
rotatable Helmholtz coil, the copper coils are at room temperature. The solenoid axis is
parallel to the ab face of the sample  and the Helmholtz coil generates a field perpendicular to the solenoid axis. Adjusting
the relative angle and the magnitude of the two fields allows us to point the
resultant field along any direction.
The Sr$_{2}$RuO$_{4}$ single crystal was grown by a floating zone
technique using an image furnace \cite{Serv05,Serv_thesis}.  Specific heat
measurements of crystals taken from the same single-crystal rod
showed volume superconductivity below a temperature of 1.31 K and
a transition width of less than 0.1K. We used 2 different samples
of plate like shape of this crystal, one having a thickness of 0.5mm with an
estimated demagnetization factor, N of 0.9 (sample 1) and the
other 0.6mm with N =0.7 (sample 2).
 The sample is cleaved along the ab -plane and AFM images show flatness down to the order of
6 $\AA$  (Fig.~\ref{Figure_1_Dresden_surface}), about twice as flat as surfaces of NbSe$_{2}$). 

\section{Coalescence, flux domains and crossing vortices}
\label{coalescence}

\begin{figure}[!]
  \includegraphics[width=9cm]{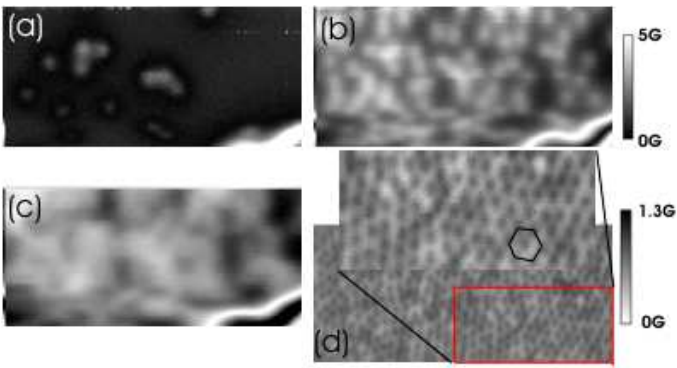}\\
  \caption{\label{Fig.2}Comparison between NbSe$_{2}$ and Sr$_{2}$RuO$_{4}$  for magnetic fields
applied perpendicular to the ab plane. (a) A $\mu$SFM image after
field cooling Sr$_{2}$RuO$_{4}$ in a field of 2 gauss, (b) in 6 gauss
 and (c) in 7 gauss. The imaging temperature is 0.35K for all
images. Coalescence of vortices is observed. (d) A $\mu$SFM image of a vortex lattice in NbSe$_{2}$. The data is
acquired after field cooling the NbSe$_{2}$ crystal in 5 gauss at a
temperature of 1.1K. The inset shows the hexagonal order of the lattice.
For all images the imaging area is 62$\mu$m $\times$30$\mu$m. }
\end{figure}

 During the imaging, the
$\mu$SQUID moved in a plane above a cleaved ab surface of the
single crystal of Sr$_{2}$RuO$_{4}$. Individual vortices
are seen \cite{Dolo05} after cooling the crystal (sample 1)  in a magnetic field
of 0.1G applied along the c-axis.  The vortices disappear completely above T = T$_{c}$ =
(1.35$\pm$0.05)K, in agreement with the T$_{c}$ value determined
previously in specific heat measurements.  At these low
fields Sr$_{2}$RuO$_{4}$ behaves as a usual type-II
superconductor.
The images of Fig.~\ref{Fig.2}
 were obtained after field cooling (FC) sample 2 in fields between 2 and 7 gauss to a temperature of 0.35K. At 2 gauss applied field, Fig.~\ref{Fig.2}(a), vortices
 are distinct,  some of them are close together, at 6 gauss Fig.~\ref{Fig.2}(b) a higher density of
 individual vortices is detected, locally  coalescing flux regions form, and  as the field
 increases further to 7 gauss,  Fig.~\ref{Fig.2}(c) the individual vortices have melted into flux domains.
 For comparison, we imaged a conventional s-wave superconductor
 NbSe$_{2}$ having a T$_{c}$ of 7.2K. NbSe$_{2}$ is a layered material, it is
 weakly anisotropic with an effective mass anisotropy ($\lambda_{c}$/$\lambda_{ab}$) of 3.3.  The
 penetration depth for applied fields along c-axis $\lambda_{ab}$ is
 0.15$\mu$m comparable with $\lambda_{ab}$ of Sr$_{2}$RuO$_{4}$.  A
 hexagonal vortex lattice is readily observed by the $\mu$SFM  Fig.~\ref{Fig.2}(d),
 after field cooling the sample in 5 gauss.  The vortices are clearly distinct
 from one another.  When the field is further increased the vortices in NbSe$_{2}$ 
 approach so close that  $\mu$SFM can't resolve the vortices anymore and the
 flux appears homogenous.
 The case of Sr$_{2}$RuO$_{4}$ is different: Instead of the formation of a vortex lattice we observe 
 vortex coalescence. The threshold value of the applied field at which the
 vortices coalesce depends on the thickness of the sample. A complete collapse of the vortices into one single domain is not
observed, probably due to the presence of weak barriers in the
material. 

\begin{figure}[t!]
  \includegraphics[width=9cm]{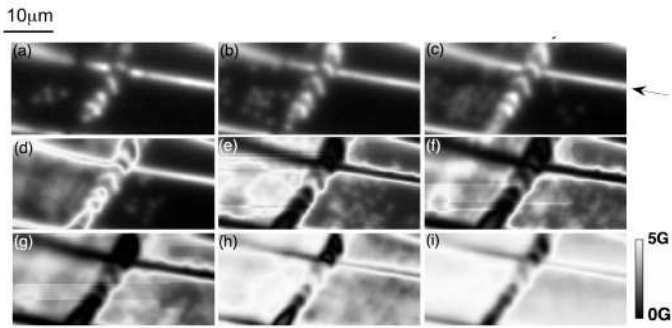}\\
  \caption{\label{Figure_3} $\mu$SFM image of 62$\mu$m $\times$30$\mu$m after
field cooling Sr$_{2}$RuO$_{4}$ in (a)2G, (b)3G, (c)4G, (d)8G, (e)9G, (f)10G, (g)11G, (h)20G, (i)50G at 0.4 K H//c.  Line defects, groves, (arrow) delimit regions of the crystal. The vortex distribution is very different between regions separated by the defects.}
\end{figure}

 Domain walls delimiting regions with angular momentum
l$_{z}=\pm1$ ($\widehat{l}$ is parallel to the c-axis) could
provide the scenario for weak intrinsic pinning at low magnetic
fields. The difference of free energy under magnetic field between the two states  \cite{Ich02,Ich05} may make  that  vortices appear preferentially in domains of one type.  This is  in qualitative agreement with our observations, as we observe flux filled and flux free regions. 
By accident we imaged an area delimited by line defects (grooves)  Fig.~\ref{Figure_3}.
We acquired images after field-cooling in fields between 2 and 50 gauss. In the normal state the magnetic flux is equally distributed throughout the sample, upon field-cooling into the superconducting state the flux in the center of the sample has to move out (Meisssner eeffect), or it stays pinned and transforms to quantized vortices. Here we observe that the region on the upper left side retains flux much more readily than the region on the other side of the line defect. Only after cooling in fields as high as 8 gauss vortices remain in the region on the right side. Different is also the high contrast along the edge of this region   Fig.~\ref{Figure_3} e-g). These observations agree with a scenario of pinning of domain walls on linear defects and thus delimiting regions with cooper pairs of different angular. This experiment indicates tentatively a way to study the order parameter domains, though a complete picture of the interaction between vortices, domain walls seperating regions of different chirality, shielding currents and surface defects has yet to be established both experimentally and theoretically.
There should be a difference between field cooled (FC) and zero-field cooled (ZFC) experiments. 
Under ZFC conditions,
domains of each chirality should be equally present. Upon
subsequent field increase the vortex should penetrate by the
domain walls and then enter from the edge preferentially in
the p- wave domain. We made ZFC experiments and increased the
applied field subsequently.  At fields less than $H_{c1}$ we
observe only a few single vortices, the shielding currents at the
surface retain the vortices, while above $H_{c1}$,  at about 30 Gauss,
vortices penetrate massively into the center of the sample. 

\begin{figure}[!]
  \includegraphics[width=9cm]{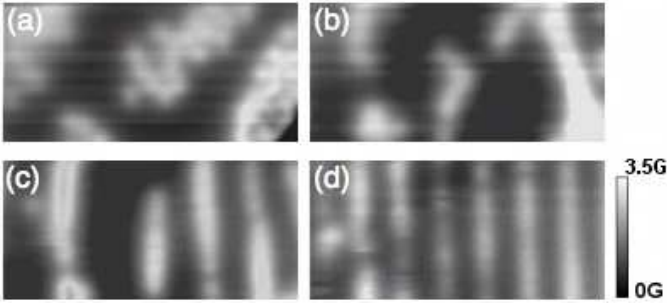}\\
  \caption{\label{Fig.4}S$\mu$SM images of flux domains in
Sr$_{2}$RuO$_{4}$ at T = 0.36K after field cooling at various fields. 
In all cases, the magnetic field amplitude  applied along c-axis (H$_{\perp}$) was kept constant at 2G while
in-plane field (H$_{ab}$) was (a) 0G, (b) 5G, (c) 10G, (d) 50G.
The imaging area is 31$\mu$m $\times$ 17$\mu$m.  Field scale
in gauss is shown on the right; dark regions are
superconducting vortex free regions.}
\end{figure}

How strongly are these domains attached to the crystal?
In order to examine the stability of the domain configuration the
in-plane field was raised while the c-axis field was kept constant.
Fig.~\ref{Fig.4} shows for increasing in-plane fields how the
condensed vortex structures rearrange freely in order to
accommodate the experimental conditions: 
For zero gauss in plane field and 2 gauss FC applied parallel with
c-axis in the sample 1 we see only domains of flux, Fig.~\ref{Fig.4}(a). The difference in flux density
between the bright (vortex) and the dark (vortex-free) regions
is 3.5 gauss. Integration of the flux pattern gives an average field
of 1.4$\pm$0.2 gauss at the $\mu$SQUID, close to the applied field of 2 gauss,
the vortices are condensed in domains, leaving entire
superconducting regions empty of flux.

At 5 gauss in-plane
applied field the flux domains become slimmer and  above 10 gauss
the flux domains are seen to evolve into line-shaped structures.
The number of flux domains was found to increase in a regular fashion with the
in-plane field amplitude. This regular increase of the
flux domain density and their temperature evolution (data not
shown) suggest that the flux structures are unrelated to any
structural defects in the crystal as defect pinning \cite{Hus92} of
vortices would interfere with regularly spaced vortex pattern.

The line-shape flux structures evoke vortex chains observed in
decoration experiments of
YBa$_{2}$Cu$_{3}$O$_{7+\delta}$ \cite{Gam92} and
Bi$_{2}$Sr$_{2}$CaCu$_{2}$O$_{8+\delta}$ \cite{Bolle91}. There vortex
chains appear when the applied field is close to the in-plane
direction of the anisotropic superconductor. Sr$_{2}$RuO$_{4}$ has
an effective mass anisotropy 
 6 times higher than
NbSe$_{2}$ and 4 times higher YBa$_{2}$Cu$_{3}$O$_{7+\delta}$ but
lower than Bi$_{2}$Sr$_{2}$CaCu$_{2}$O$_{8+\delta}$. Therefore
the arrangement of the domains in lines may be driven by
anisotropy.
The attraction between vortices in anisotropic superconductors comes from the misalignment between the vortex axis (B) and the direction of the applied field (H)  giving raise to a net transverse magnetization M. 
This attractive interaction \cite{Buz90}
between the vortices is directed along the plane spanned by the anisotropy axis
and the in-plane applied field. We observe also that the linear pattern follow the direction of the applied field
when the field is rotated in the ab plane (images not shown), this proofs that the coalescence of the vortices in domains does not
originate from defects as the magnetic energies in play can align
these domains as vortex chains.

\begin{figure}[!]
 \includegraphics[width=9cm]{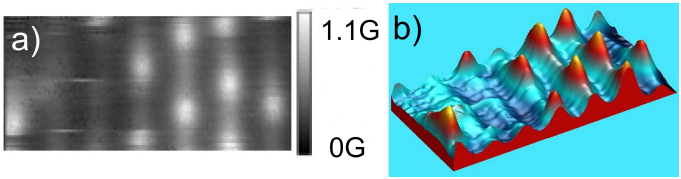}\\
 \caption{\label{Fig.5}Crossing vortices in Sr$_{2}$RuO$_{4}$.
Figure (a) 10 G FC and increased to 50 G at low T. The magnetic scale is shown on the right. The dimensions are 31$\mu$m$\times$15$\mu$m and the imaging temperature is 0.35K. The tilting angle is 87$^{\circ}$. The (b) image is the 3D representation of the image (a).}
 \end{figure}

In isotropic superconductors at low tilted fields the flux lines
penetrate parallel to each other and with the average field in the
sample and  arrange in a pattern that minimizes the interaction
energy. If the superconductor is strongly anisotropic the free
energy of the vortex state has two local minima and consequently a
flux line can penetrate in two distinct directions in the material.
For high anisotropy only flux lines that are nearly parallel or perpendicular to the layers are stable
\cite{Sud93}. This model predicted that two types of vortices may be observed in Sr$_{2}$RuO$_{4}$, formed by two interpenetrating lattices, a lattice of vortices parallel to the layers and crossing vortices that are nearly parallel to the c-axis.
We undertook measurements changing
the applied magnetic field while the sample is in the
superconducting state Fig.~\ref{Fig.5}. The magnetic field component parallel to the
plane is higher than the first critical field H$_{c1}$  in the in-plane direction (H$_{c1}^{ab}$=10G), while the perpendicular component is lower than H$_{c1}$(H$_{c1}^{c}$=50G). We observe crossing vortices decorating flux channels Fig.~\ref{Fig.5}. 

 \section{Conclusion}
 \label{conclusion}
We observe vortex coalescence and spatially dependent flux pinning, consistent with what might be expected from a chiral order parameter in Sr$_{2}$RuO$_{4}$.  Tilting the magnetic field, anisotropy effects  become
important so the vortices try to align along the field direction, forming vortex chains. We observe the simultaneous presence of two vortex orientations in the crystal, one long the planes, flux channels, and one perpendicular to it, attributed to the decoration of in-plane flux channels by Abrikosov vortices.

\section{Acknowledgments}
 We thank O. Fruchard for his help in acquiring the AFM images of the crystal surfaces. 
 
\bibliographystyle{elsart-num}
\bibliography{XBib_Dresden}







\end{document}